%% file: main.tex
\documentclass[letterpaper, paper,11pt]{AAS}		

\usepackage{bm}
\usepackage{amsmath}
\usepackage{subfigure}
\usepackage{cite}
\usepackage{footnpag}			      	

\usepackage{amssymb}  
\usepackage{amsthm}

\usepackage{subfiles}

\PaperNumber{18-253}

\begin{document}

\title{Robust Myopic Control for Systems with Imperfect Observations}

\author{Dantong Ge\thanks{Ph.D. Candidate, Department of Aerospace Engineering, Beijing Institute of Technology, Beijing, China 100081; Visiting Student, Department of Aerospace Engineering and Engineering Mechanics, University of Texas at Austin, Austin, TX 78712},
\ Melkior Ornik\thanks{Postdoctoral Fellow, Institute for Computational Engineering and Sciences, University of Texas at Austin, Austin, TX 78712},
\ and Ufuk Topcu\thanks{Assistant Professor, Department of Aerospace Engineering and Engineering Mechanics; Institute for Computational Engineering and Sciences, University of Texas at Austin, Austin, TX 78712}
}

\maketitle{}

\begin{abstract}
Control of systems operating in unexplored environments is challenging due to lack of complete model knowledge. Additionally, under measurement noises, data collected from onboard sensors are of limited accuracy. This paper considers imperfect state observations in developing a control strategy for systems moving in unknown environments. First, we include hard constraints in the problem for safety concerns. Given the observed states, the robust myopic control approach learns local dynamics, explores all possible trajectories within the observation error bound, and computes the optimal control action using robust optimization. Finally, we validate the method in an OSIRIS-REx-based asteroid landing scenario.

\end{abstract}

\section{Introduction}
\label{Introduction}
\input{Introduction.tex}

\section{Preliminaries}
\label{Preliminaries}
\input{Preliminaries.tex}

\section{Problem Statement} 
\label{ProblemDefinition}
\input{ProblemStatement.tex}

\section{Myopic Control with Hard Safety Constraints}
\label{StochasticLearning}
\input{MC.tex}

\section{Robust Myopic Control}
\label{RobustLearning}
\input{RMC.tex}

\section{Simulation}
\label{SimulationResults}
\input{Simulation.tex}

\section{Conclusion}
\label{Conclusion}
\input{Conclusion.tex}

\section{Acknowledgment}
This research was funded by grants NSF1617639 and NSF1646522 from the National Science Foundation. The authors thank Srilakshmi Pattabiraman for offering helpful ideas in formulating the initial problem.

\appendix
\section*{Appendix: Proof of Theorem 1}
\label{Appendix}
\input{Appendix.tex}

\bibliographystyle{AAS_publication}   
\bibliography{main}   

\end{document}

%% file: Introduction.tex
An accurate dynamics model is indispensable in developing control strategies for systems moving in a complex and changing environment. When exploring a new environment, however, prior knowledge of the dynamics model is usually scarce \cite{sutton1998reinforcement}. Consider a small body landing mission. System dynamics are complex due to the object's irregular shape, non-uniform mass distribution, weak gravitational field, as well as environmental perturbations caused by solar pressure and gravity of other celestial body \cite{ulamec2009surface,gal2015osiris}. Without long-term close observation, a precise landing dynamics model is usually unavailable before launch \cite{berry2013osiris}. To decrease the impact of model uncertainty on the actual system behavior, the system should update its onboard model and make real-time decisions based on interactions with the environment.

Existing methods for data-driven learning \cite{brunton2016discovering,hills2015algorithm}, control design and performance assessment under unknown dynamics \cite{khadraoui2016adaptive,ahmadi2017safety} typically require intensive measurement data or significant information about the model, such as pre-designed controllers under different conditions, parameter variation range, or state characteristics. Nevertheless, in the problem setting studied in this paper, state measurements are limited. Furthermore, we assume that there is no prior knowledge of the environment. In \cite{faust2014continuous}, a sampling-based task learning method is proposed for control-affine systems with unknown dynamics. The method attains the optimal control policy through a sequential learning of a global state-value function in the state space and a local action-value function around each state. Although the method is independent of actual models, it cannot be applied in asteroid landings, since exploring the entire state space in a limited time is impractical. 

The concept of myopic control was first proposed in \cite{Ornik}, where the method learns local dynamics by observing the variation of states under a given control sequence, and uses the estimated dynamics to make online decisions. A goodness function is developed to quantify the system performance with respect to mission objectives, and the control action that yields the maximal value of the goodness function is selected as the piecewise constant control for the next step. The ability of learning dynamics and computing control actions in real-time makes myopic control a good fit to be implemented in our problem, except that, in the method, accurate state observations are assumed to be available at every sampling point. In practice, observation information obtained from sensors is usually of limited precision due to measurement noise or instrument calibration error \cite{singla2006adaptive}. Such imperfect state observations may be misleading in decision-making as, for example, the generated control action may guide the vehicle to an unexpected direction. Another drawback of myopic control is that hard constraints are not guaranteed to be satisfied in the context. Although the method assigns a small value to the goodness function for undesirable system performance so that the corresponding control action is less likely to be selected, it does not exclude the possibility of constraint violation. As hard constraints define the feasible regions in the state space, it is required for safety concern that they are strictly satisfied all the time.


In this paper, we extend the work of myopic control with hard safety constraint guarantees, and propose a robust myopic control strategy in the presence of imperfect observations. Instead of optimizing based on a single observed trajectory, the method considers the set of all possible trajectories within a given observation error bound, and computes the optimal control actions using robust optimization. 

We validate the approach in the setting of an on-going asteroid mission OSIRIS-REx with landing requirements \cite{gal2015osiris}, and compare the system performance of both nominal myopic control and robust myopic control under the same observation conditions.


%% file: Preliminaries.tex
In this section, we give a brief overview of the notation, necessary assumptions, and the general idea of myopic control. We denote the set $\{0,1,\dots,m \}$ by $[m]$ and the set of all continuous functions from set $\mathcal{A}\subseteq \mathbb{R}^m$ to set $\mathcal{B}\subseteq \mathbb{R}^n$ by $C(\mathcal{A}, \mathcal{B})$. For a vector $v\in \mathbb{R}^n$, $\Vert v\Vert$ denotes its 2-norm. For $x\in \mathbb{R}^n$, $x_i$ with $i\in {1,2,...,n}$ denotes the $i$-th component of $x$. Notation $f|_{[a,b]}$ emphasises that function $f$ is only considered on the interval $[a,b]$. Symbol $\mathbb{N}$ denotes all strictly positive integers.

Consider the system with compact domain $\mathcal{X} \subseteq \mathbb{R}^n$ 
\begin{align}
\dot{x} = f(x) + \sum_{i=1}^{m} g_i(x)u_i \label{dyn}
\end{align}
where functions $f, \{g_i\}_{i \in [m]} \in C(\mathcal{X}, \mathcal{X})$ and control input $u=[u_1,...,u_m]^T\in \mathbb{R}^m$. Let the solution of \eqref{dyn} under control signal $u:[0,\infty)\to\mathbb{R}^m$ and initial state $x_0$ be denoted by $\phi_{u}(\cdot,x_0)$. We assume that $u$ is piecewise-constant and a unique solution $\phi_{u}(\cdot,x_0)$ under $u$ exists.

We make the following assumptions:
\begin{itemize}
\item No prior knowledge of the dynamics is available except it is in the form of \eqref{dyn}. Functions $f$ and $\{g_i\}_{i \in [m]}$ are unknown, but are known to be bounded by $M_0$ and are $M_1$-Lipschitz, i.e., for some $M_0>0,M_1> 0$,
\begin{equation}
\begin{split}
\Vert f(x) \Vert &\leq M_0, \\
\Vert g_i(x) \Vert &\leq M_0, \forall i \in [m], \\
\Vert f(x) - f(y) \Vert &\leq M_1 \Vert x-y \Vert, \forall x,y \in \mathcal{X}, \\
\Vert g_i(x) - g_i(y) \Vert &\leq M_1 \Vert x-y \Vert, \forall x,y \in \mathcal{X}, \forall i \in [m].
\end{split}
\end{equation}  
\item The system starts at an initial state $x_0$ and runs only once. All control decisions are made during the run with no repetitions.
\end{itemize}

For any given control action $u$, a \textit{goodness function} is defined as an evaluation of how well the system performs according to the control objective
\begin{align}
(\phi, v) \mapsto G(\phi, v),  \phi \in \mathcal{F}, v \in \mathbb{R}^n
\end{align}
where
\begin{align}
\mathcal{F} = \bigcup_{T \geq 0} C([0,T], \mathcal{X})
\end{align}
The system performance to be evaluated includes the trajectory $\phi$  until time $T$ and the system velocity $v$ at time $T$. The Lipschitz constant $L$ of function $G:\mathcal{F}\times\mathbb{R}^n \rightarrow \mathbb{R}$ for all $\phi_1|_{[0,T_1]}, \phi_2|_{[0,T_2]} \in \mathcal{F}$ and $v_1, v_2 \in \mathbb{R}^n$ is given by
\begin{align}
\label{lip1}
|G(\phi_1|_{[0,T_1]}, v_1)-G(\phi_2|_{[0,T_2]}, v_2)| \leq L(d(\phi_1|_{[0,T_1]}, \phi_2|_{[0,T_2]})+\Vert v_1-v_2\Vert),
\end{align}
where
\begin{align}
\label{lip2}
d(\phi_1|_{[0,T_1]}, \phi_2|_{[0,T_2]})=|T_1-T_2|+\max_{t\in [0, \min(T_1, T_2)]}\Vert \phi_1(t)-\phi_2(t)\Vert.
\end{align}

Generally, the goodness function predicts how different control actions will influence the system behavior in the future, and the one that maximizes the goodness function is regarded as the optimal control action. Thus, the problem to be solved can be addressed as finding a control strategy for the system that always results in the maximum goodness function value, i.e., finding a control signal $u^* : [0,T] \rightarrow \mathcal{U}$ such that, for all $t\in [0,T]$, if $x=\phi_{u^*}(t,x_0)$, then
\begin{align}
\label{eqw2}
 G(\phi_{u^*}(\cdot,x_0)|_{[0,t]},f(x)+\sum_{i=1}^m g_i(x)u^*_i(t))=\max_{u \in \mathcal{U}} G(\phi_u(\cdot,x_0)|_{[0,t]},f(x)+\sum_{i=1}^m g_i(x)u_i).
\end{align}

We now give a brief description of myopic control and refer the reader to \cite{Ornik} for details of an approximate solution to \eqref{eqw2}. The myopic control strategy is built upon sequential and repeating phases of learning local dynamics and computing optimal control actions. First, we design a goodness function $G(\phi,v)$ according to the control objective. In order to learn the dynamics, $m+1$ affinely independent control inputs are applied on the system for a given short period of time and the state variations are observed. Assume that the affinely independent control inputs are marked as $u^*+\Delta u^0,...,u^*+\Delta u^m$ with $u^*=0$ at time $t_0$. By applying  $u^*+\Delta u^j, j=0,...,m$ in $[t_0+j\varepsilon,t_0+(j+1)\varepsilon]$, the system state at the end of the time interval is observed and marked as $x_{j+1}$. The corresponding trajectory composed by these states is denoted as $\phi$. The local dynamics are then approximated by 
\begin{equation}
\label{definethisdantong}
v(u, x)=\sum^m_{j=0} \lambda_j(x_{j+1}-x_j)/\varepsilon,
\end{equation}
where $\lambda_0,...,\lambda_m$ are unique coefficients determined by $\sum \lambda_j=1$ and $u=\sum\lambda_j(u^*+\Delta u^j)$. The optimal control action that maximizes the goodness function is obtained as 
\begin{equation}
\label{dantong2}
u^* \in \text{arg}\max_{u\in\mathcal{U}} G(\phi(\cdot,x_0)|_{[0,t_0+(m+1)\varepsilon]},v(u,x))
\end{equation}
Such a $u^*$ is then used as the basic control input in the next iteration, with the new $t_0$ given by $t_0+(m+1)\varepsilon$. For simplicity, we denote the entire control signal derived from this strategy as $u^*:[0,T]\to\mathcal{U}$. The process repeats until task completion.

%% file: ProblemStatement.tex
In this paper, we improve the basic method for myopic control in \cite{Ornik} by accounting for the existence of hard safety constraints \cite{schouwenaars2004receding} and availability of only imperfect system state observations \cite{van2011lqg}. We define \emph{hard safety constraints} as safety-related state constraints that need to be strictly satisfied. By denoting the set of all hazardous states as $\mathcal{B}\subseteq \mathcal{X}$, the states that satisfy safety constraints are given by $x(t)\in\mathcal{X}\backslash\mathcal{B}$. Besides, the observed states attained from sensors often deviate from the true ones due to measurement noise and instrumental error \cite{kirkup2006introduction}. We consider such \textit{imperfect} observations in the problem, and use $x^{\text{obs}}$ to distinguish the observed state from the true state $x^\text{true}$. We assume that even though state observations are inaccurate, there is a known constant $\Delta>0$ such that $\Vert x^\text{true}(t)-x^\text{obs}(t)\Vert \leq \Delta$ for all $t\geq 0$. The problem to be solved in this paper can be described as

\textbf{Problem 1} 
\emph{Let the initial state $x_0 \in \mathcal{X}$, the unsafe set $\mathcal{B} \subseteq \mathcal{X}$, and the feasible control action set $\mathcal{U}\subseteq \mathbb{R}^m$ be given. Additionally, let $G$ be the goodness function designed from control objective, $T\geq0$ be the time length of the mission, and $\Delta$ be the bound of state observation error. Find a control signal $u^* : [0,T] \rightarrow \mathcal{U}$ that satisfies the following conditions for all $t\in [0,T]$:\\
(i) $\phi_{u^*}(t,x_0)\not\in \mathcal{B}$,\\
(ii) $G(\phi_{u^*}(\cdot,x_0)|_{[0,t]},v(u^*,\phi_{u^*}(t)))=\max_{u \in \mathcal{U}} G(\phi_u(\cdot,x_0)|_{[0,t]},v(u,\phi_u(t)))$.}

In the remainder of the paper, we first propose a modified myopic control strategy to provide guarantees for hard safety constraints. Then we solve the above problem with imperfect observations by developing a robust myopic control strategy.

%% file: MC.tex
In myopic control, the goodness function predicts system behavior in the near future based on the learned local dynamics. As a result, it can also foresee if there is any conflict between system safety and mission objective, i.e., whether the system will violate hard safety constraints when it is trying to achieve the goal. In this paper, we estimate system performance under different control actions at the beginning of every learning period. For control inputs that lead to constraint violation, we set their goodness function values as $-\infty$. Otherwise, the goodness function receives the same real value as designed before. We use this method of assigning goodness function values so that by maximizing the goodness function, the control action that causes constraint violation will not be chosen, provided there are any better choices. In the following, we present two possible ways for system performance prediction:\\
(i) Assume that the dynamics remain the same for a short time interval. We can directly predict system state $\phi_{u}^\text{pred}(t+\Delta t,x_0)$ under every possible control action $u$ for some small $\Delta t$ (e.g. a learning period). We regard a control action as \emph{bad} if the predicted state belongs to set $\mathcal{B}$. Then we set its goodness function value as $-\infty$, i.e., if $$\phi_{u}^\text{pred}(t+\Delta t,x_0) \in \mathcal{B},$$ then $$G(\phi_u(\cdot,x_0)|_{[0,t]},v(u,x))= -\infty.$$

(ii) Assume that the dynamics keep changing, but remain Lipschitz-continuous with a known Lipschitz constant. Then we can compute the set $\Phi_u^\text{pred}(t+\Delta t,x_0)$ of all possible values for $\phi_{u}(t+\Delta t,x_0)$. We regard a control action as \emph{bad} if any element in the set of predicted system state from time $t$ to $t+\Delta t$ is inside $\mathcal{B}$, i.e. if, using the given control action, there is possible that the system will violate the state constraint. We then set the goodness function value as $-\infty$, i.e., if $$\bigcup_{t'\in[t,t+\Delta t]}\Phi_{u}^\text{pred}(t',x_0) \cap \mathcal{B} \neq \emptyset,$$ then $$G(\phi_u(\cdot,x_0)|_{[0,t]},v(u,x))= -\infty.$$

\begin{figure}[h]
\centering
\includegraphics[width=0.4\textwidth]{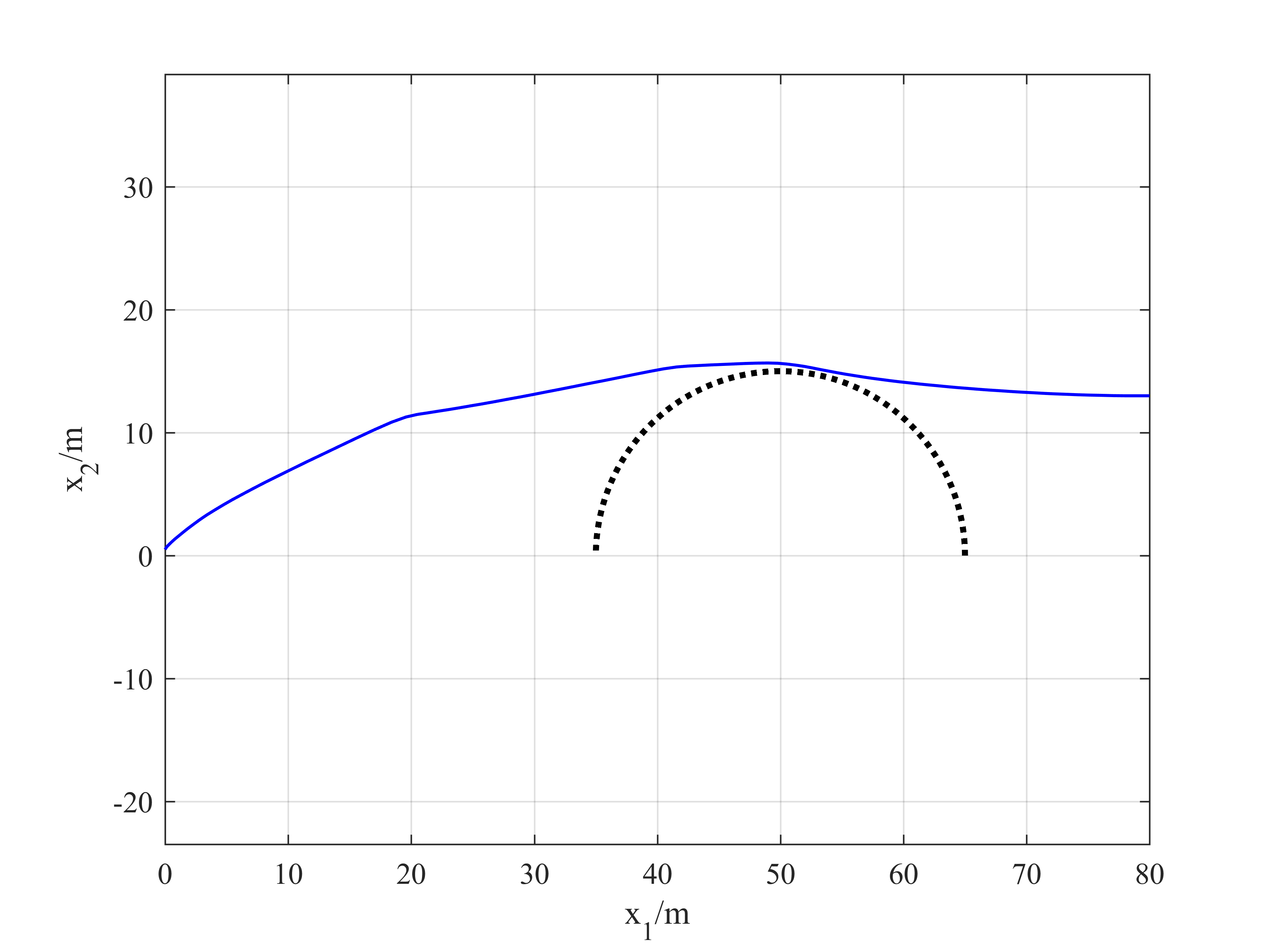}
\caption{\label{fig:s1} An illustration of the system trajectory, in the $x_1$-$x_2$ plane, in \emph{Example 1}. The obstacle is denoted by a black dashed semicircle, and the trajectory is given in solid blue.}
\end{figure}

To demonstrate the effect of such goodness function design, we present a simple example where the system is required to avoid an obstacle:\\
\emph{Example 1} Consider an agent moving with the dynamics
\begin{align}
\label{damdam}
 \dot{x}_1=x_3,\quad \dot{x}_2=x_4,\nonumber\\
 \dot{x}_3=u_1,\quad \dot{x}_4=u_2.
\end{align}
Equation \eqref{damdam} describes the movement of an agent in $\mathbb{R}^2$, where $x_1$ and $x_2$ are the agent's position, and $x_3$ and $x_4$ are its velocity. Note that in the example, we only use the dynamics model to generate state observations and use no knowledge of it in determining control inputs. We further define the obstacle region as $\mathcal{B}=\{x\in\mathbb{R}^4~|~(x_1-50)^2+x_2^2\leq 15^2, x_2\geq 0 \}$ and design the goodness function in the following form
\begin{eqnarray}G(\phi_u, v(u, x))=
\begin{cases}
-\Vert r_p - r_f \Vert^2-\frac{\tau}{\Vert r_p - r_\mathcal{B} \Vert^2}, &r_p \not\in \mathcal{B}\cr -\infty, r_p \in \mathcal{B}\end{cases}
\end{eqnarray}
where $r_f=[0,0]^T$ is the target position for $(x_1,x_2)$, and $r_p$ is the predicted position of $(x_1,x_2)$ under the assumption of (i). The generated trajectory with $\tau=150$ is shown in Figure \ref{fig:s1}, where the system successfully avoids the obstacle and reaches the target. 

Having introduced a goodness function design with guaranteed hard constraint satisfaction, we now turn our attention to the existence of imperfect system state observations.

%% file: RMC.tex
\begin{figure}[h]
\centering
\includegraphics[width=0.4\textwidth]{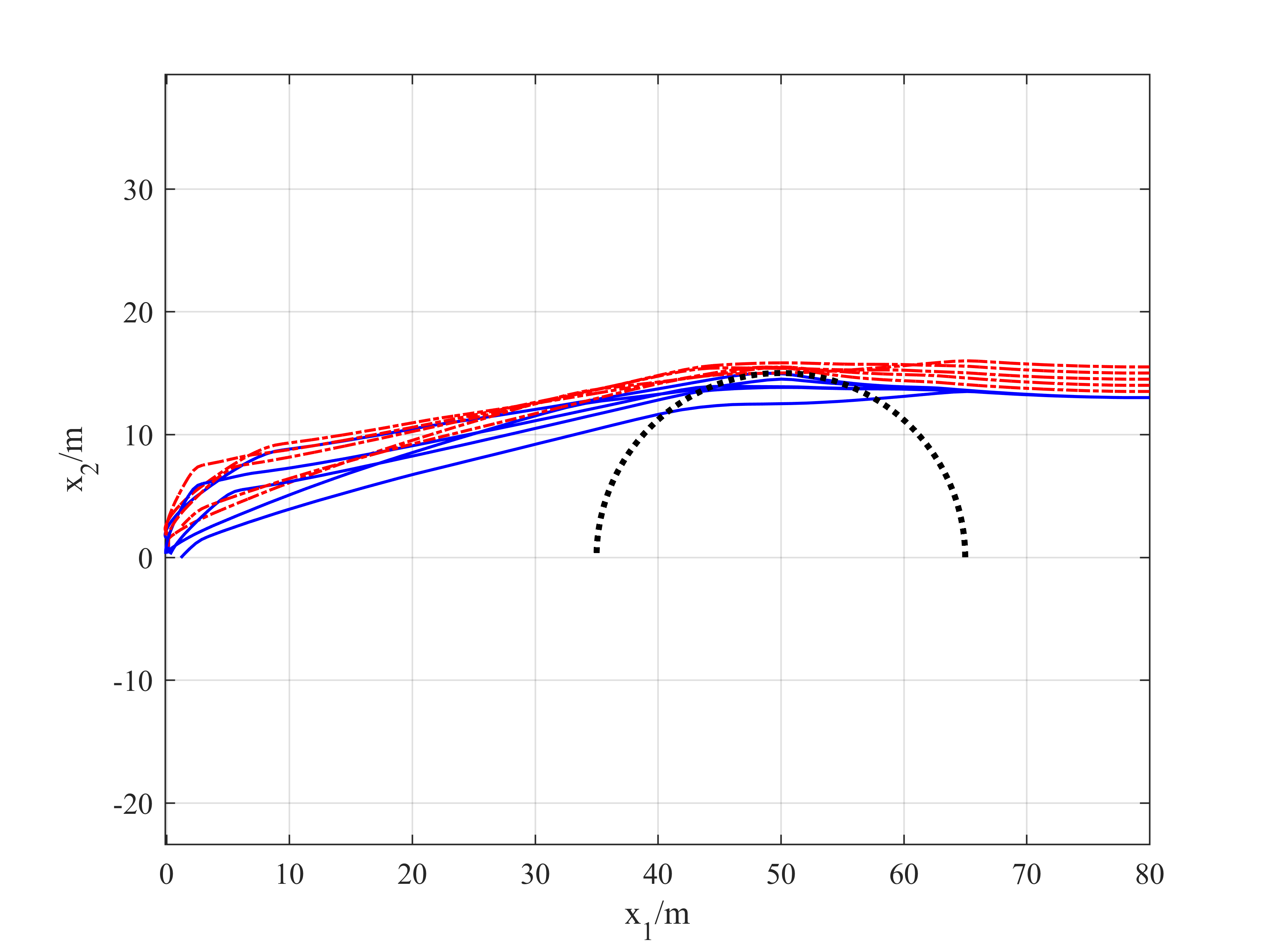}
\includegraphics[width=0.4\textwidth]{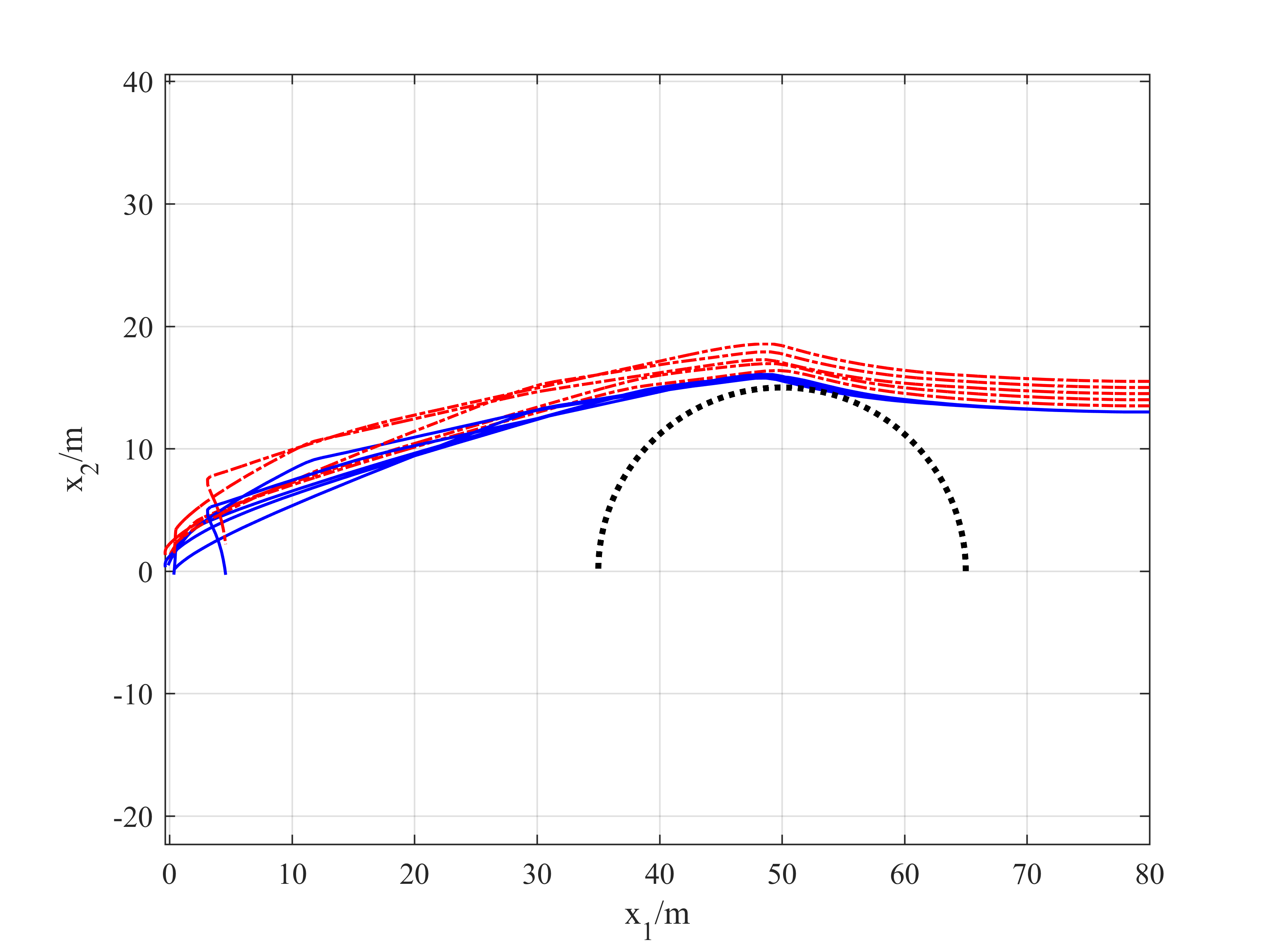}
\caption{\label{fig:s2} Trajectories generated by nominal myopic control (left) and robust myopic control (right) under constant observation errors $e=[0.5, 1.0, 1.5, 2.0, 2.5]$. Blue solid lines refer to the actual trajectories, and red dashed lines refer to the observed trajectories.}
\end{figure}

In the presence of imperfect state observations, the myopic control method will learn an incorrect local dynamics and generate control actions that potentially lead the system to an unexpected direction. We illustrate this phenomenon in the previous example, by adding 5 different constant observation errors on $x_2$ i.e., $x^\text{obs}_2=x^\text{true}_2+e$. The obtained trajectories are depicted on the left side of Figure \ref{fig:s2}. As can be seen, all the actual trajectories collide with the obstacle because the system wrongly believes that it is further away.

To overcome the constraint violation problem under imperfect observations, we propose a robust myopic control strategy that ensures tolerable performance for all possible system trajectories. Assume that the observation error bound $\Delta$ is known, i.e., the actual system state $x^\text{true}(t)$ satisfies $x^\text{true}(t)\in[x^{\text{obs}}(t)-\Delta,x^{\text{obs}}(t)+\Delta]$ for all $t\in[0,T]$. Notice that there exist multiple possible trajectories consistent with the same set of observations. Using the local dynamics learned from the observed states, we predict possible system performance under different control actions for each trajectory. Analogous to classic robust control algorithms which take bounded modeling errors into account and optimize system performance by solving a min-max problem \cite{bemporad2003min,campo1987robust}, we design the robust myopic control strategy. 
\begin{figure}[h]
\centering
\includegraphics[width=0.4\textwidth]{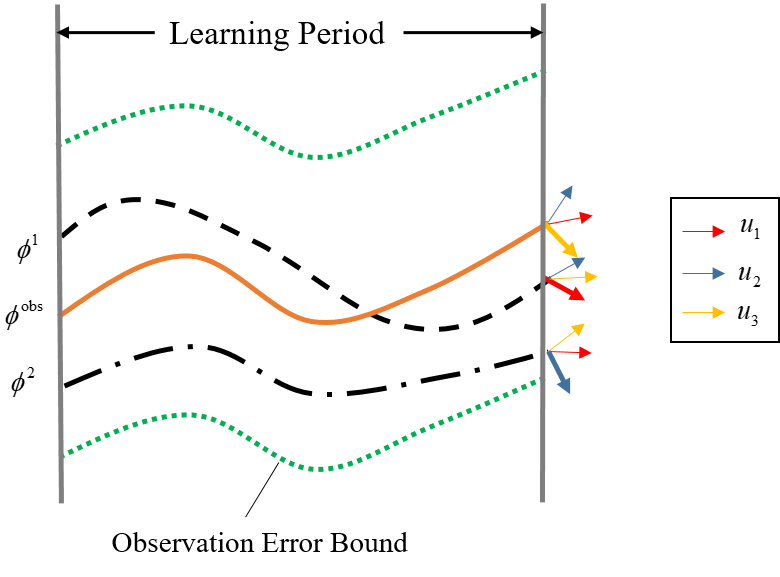}
\caption{\label{fig:robillu} An illustration of the robust myopic control strategy. The observed system trajectory is given in red, and two additional possible system trajectories consistent with known observation error bounds are given in black. At the end of the learning period, we consider each of the three possible controls ($u_1$, $u_2$, and $u_3$), and calculate the predicted system velocity for each of those controls, at each possible trajectory. The goodness function is designed to be proportional to the upwards slope of the predicted velocity vectors. As the worst slope generated by $u_1$ for different possible trajectories is better than the worst slopes of $u_2$ and $u_3$, the robust control algorithm will choose $u_1$ as its control input for the next learning period.}
\end{figure}
As illustrated in Figure \ref{fig:robillu}, for every control input $u$, the strategy finds its ``worst case'', i.e., a trajectory that results in the worst performance when $u$ is applied. It then searches for the control action $u^+$ whose worst performance is the best among others. 
Formally, instead of the control input $u^*$ from \eqref{dantong2}, the robust myopic control strategy finds a control input $u^+ \in \mathcal{U}$ at time $t_0+(m+1)\varepsilon$ such that
\begin{align}
\label{dandan}
 G(\phi,v^\phi(u^+,x))=\max_{u \in \mathcal{U}} \min_{\phi\in\mathbb{B}(\phi^\text{obs},\Delta)} G(\phi,v^\phi(u,x)),
\end{align}
where $v^\phi$ is the approximate system direction defined in \eqref{definethisdantong}, given the trajectory $\phi:[0,t_0+(m+1)\varepsilon]\to\mathbb{R}^n$, $\phi^\text{obs}$ is the observed system trajectory, and $\mathbb{B}(\phi^\text{obs},\Delta)=\{\phi:[0,t_0+(m+1)\varepsilon]\to\mathbb{R}^n~|~\|\phi(t) - \phi^{ \text{obs}}(t)\| \leq \Delta\text{ for all } t\in[0,t_0+(m+1)\varepsilon]\}$. The control input $u^+$ is then applied after modified by $\Delta u^j$, as described in the preliminary section. We assume that the $\min$ in \eqref{dandan} exists, which holds under mild assumptions of continuity of $G$. Alternatively, a nearly-robust optimal control action $u^+$ can be obtained by choosing $\phi$ such that $G(\phi,v^\phi(u,x))$ is close to $\sup_{\phi\in\mathbb{B}(\phi^\text{obs},\Delta)} G(\phi,v^\phi(u,x))$. For simplicity, as in the preliminary section, we slightly abuse notation and in the future also use $u^+$ to denote the control signal computed from the robust myopic control method.

We now provide a theoretical bound for system performance under robust myopic control $u^+$. We write $\phi^\text{true}_{u^+}$ for the true trajectory under control signal $u^+$, and use $v^\text{true}(u,x)= f(x) + \sum_{i=1}^{m} g_i(x)u_i$. We want to determine the level of suboptimality of the robust myopic control law, i.e., bound
\begin{align}\label{tobound}|G(\phi^\text{true}_{u^+}|_{[0,t]}, v^\text{true}(u^+(t),\phi^\text{true}_{u^+}(t)))-\max_{u\in \mathcal{U}}G(\phi^\text{true}_{u^+}|_{[0,t]}, v^\text{true}(u(t),\phi^\text{true}_{u^+}(t)))|.
\end{align}
In the following, we derive the theoretical bound based on the results in \cite{Ornik}, which only covers the case where the goodness function is real-valued and has a Lipschitz constant in the sense of \eqref{lip1}--\eqref{lip2}. In the case where the goodness function can reach $-\infty$, it can be easily shown that the robust control law will always choose a control action that results in real-valued goodness, if such a control action exists. Thus, we have

\textbf{Theorem 1} 
\textit{Let functions $f, \{g_i\}_{i \in [m]}$ in system dynamics be bounded by $M_0$ and be $M_1$-Lipschitz. Assume that the goodness function $G$ is real-valued and has Lipschitz constant $L$, and that observation noises are bounded by $\Delta$. Then,}
\begin{equation}
\begin{split}
&|G(\phi^\text{true}_{u^+}|_{[0,t]}, v^\text{true}(u^+(t),\phi^\text{true}_{u^+}(t)))-\max_{u\in\mathcal{U}} G(\phi^\text{true}_{u^+}|_{[0,t]}, v^\text{true}(u,\phi^\text{true}_{u^+}(t)))| \\
& \leq 8L M_0 M_1 (m+1)^3 (4 m^{\frac{3}{2}} + \delta) \frac{\varepsilon}{\delta}+2L\left(3\Delta+ \frac{4\Delta}{\varepsilon}\left( 1 + 4m \frac{\sqrt{m}}{\delta}\right)\right)
\end{split}
\end{equation}
\textit{for all $t=k (m+1)\varepsilon$, $k\in\mathbb{N}$.}

We give the proof of Theorem 1 in Appendix. 

\textbf{Remark 1}
\textit{While Theorem 1 provides bound only for $t=k(m+1)\varepsilon$, i.e., end times of each learning cycle, a similar bound can be obtained for all $t\geq(m+1)\varepsilon$, in analogy to the results in \cite{Ornik}. Such a result is a consequence of Lipschitz continuity of the system dynamics, as well as the bound on the size of test controls $\Delta u^j$.}

\textbf{Remark 2} 
\textit{We note that the bound developed in Theorem 1 is extremely liberal, and can likely be reduced by employing a more careful, and technical, proof procedure. Nonetheless, the resulting bound shows that parameters $\delta,\varepsilon$ and distribution of the observation noises have a joint influence on the expected system performance, and that by reducing observation noise and properly selecting parameters, robust myopic control achieves performance arbitrarily close to optimal.}


We now apply the robust myopic control strategy to \textit{Example 1}. The same five constant observation errors are added on $x_2$, and the results are shown on the right side of Figure \ref{fig:s2}. In the simulation, we set the observation error bound $\Delta$ as 1.5 times larger than the actual errors. The right side of Figure \ref{fig:s2} shows that all system trajectories now avoid the obstacle.

%% file: Simulation.tex
\subsection{Asteroid Model}
In this section, we apply the proposed method to an asteroid landing scenario based on the ongoing OSIRIS-REx asteroid sample return mission \cite{gal2015osiris}. Launched in 2016, the spacecraft is expected to arrive at the asteroid 1999 RQ36 Bennu in late 2018. After performing close-proximity mapping operations for about 12 months, the spacecraft is going to descend towards the surface of the planet for sample collection. Currently, a rough asteroid model and its associated gravity field model are available based on ground observations \cite{nolan2013}. In the simulation, we use the model of dynamics described in \cite{furfaro2013} to generate the actual states of the system:
\begin{equation}
\label{moddyn}
\begin{split}
\dot{r}_x&=v_x,\\
\dot{r}_y&=v_y,\\
\dot{r}_z&=v_z,\\
\dot{v}_x&=2\omega v_y+\omega ^2r_x+\frac{\partial V}{\partial r_x}+u_x+p_x,\\
\dot{v}_y&=-2\omega v_x+\omega ^2r_y+\frac{\partial V}{\partial r_y}+u_y+p_y,\\
\dot{v}_z&=\frac{\partial V}{\partial r_z}+u_z+p_z,
\end{split}
\end{equation}
where $\boldsymbol r=[r_x,r_y,r_z]^T$ is the lander position, $\boldsymbol v=[v_x,v_y,v_z]^T$ is the lander velocity, and $\boldsymbol u=[u_x,u_y,u_z]^T$ is the control vector corresponding to engine thrust, all given in the body-fixed rotating frame. Additionally, $\omega $ is the asteroid rotating speed, $V$ is the potential field of the asteroid, and $\boldsymbol p=[p_x,p_y,p_z]^T$ is the environmental perturbation caused by solar pressure and gravitation of other celestial body. For simulation purpose, we simply set $\boldsymbol p=\boldsymbol 0$ and directly use the potential field model developed in \cite{nolan2013}. We omit the details of this model, as it is complex and does not play a large role in the remainder of the paper. We emphasize that, in line with the setting of myopic control, no specific knowledge of the above dynamics model is used in computing control actions --- we simply use model \eqref{moddyn} to simulate system behavior.

\subsection{Goodness Function}
To achieve high landing accuracy, the distance between the target and the predicted touchdown location should be as small as possible. We denote by $r_f$ the position of the designated landing site, and $r_p(t')$ the predicted system state at time $t'$. In our prediction, we assume that the vehicle, if its current position, at time $t$, is $r$ and velocity $v$, flies with a constant acceleration $a_u$ during $[t,t']$, i.e.,
\begin{align}
\label{pred}
r_p(t') =r+(t'-t)v+\frac{1}{2}(t'-t)^2a_u.
\end{align}
where $a_u$ is the predicted system acceleration when control input $u$ is applied at time $t$.

The landing error is then calculated from $\Vert r_p(t_{go})-r_f\Vert$, where $t_{go}$ is the estimated landing time, i.e., time that $r_p$ intersects with the asteroid surface. As we seek to minimize this error, we define the goodness function as
\begin{align}\label{goodfunc}
G(\phi|_{[0,t]}, v(u, \phi(t)))&=-\Vert r_p(t_{go}) - r_f \Vert^2,
\end{align}
It is possible that the predicted trajectory $r_p$ keeps extending in the space and never intersects with the asteroid surface. In such a case, we set a flight time upper bound $t_\text{max}$, and define the ``landing error'' as the minimum distance between the lander and the asteroid in the prediction. Thus, we amend \eqref{goodfunc} by
\begin{align}
G(\phi|_{[0,t]}, v(u, \phi(t)))&=
\begin{cases}
-\Vert r_p(t_{go}) - r_f \Vert^2,\quad t\leq t_{go}\leq t_\text{max},\\
-\min_{t''\in[t, t_{go}]}\Vert r_p(t'') - r_f \Vert^2,\quad t_{go}>t_\text{max}.
\end{cases}
\end{align}

Due to resolution constraints, the current best available map of Bennu can only model the asteroid as largely convex and locally flat. In order to showcase the quality of the proposed landing approach, we add an additional dangerous obstacle on the asteroid surface and place it close to the landing site. The additional obstacle poses a hard safety constraint to the problem, as landing on it is prohibited. We define the hazardous states as $\mathcal{B}=\{r|(r-r_\mathcal{B})^2\leq R\}$, where $r_\mathcal{B}$ is the center point of the semi-spherical obstacle. Upon consideration of hard constraints, we modify the goodness function as
\begin{equation}
\label{defgood}
G(\phi|_{[0,t]}, v(u, \phi(t)))=
\begin{cases}
-\Vert r_p(t_{go}) - r_f \Vert^2-\frac{\tau}{\Vert r_p(t') - r_\mathcal{B} \Vert^2}, \quad r_p(t') \not\in \mathcal{B}, t\leq t_{go}\leq t_\text{max},\cr
-\min_{t''\in[t, t_{go}]}\Vert r_p(t'') - r_f \Vert^2-\frac{\tau}{\Vert r_p(t') - r_\mathcal{B} \Vert^2}, \quad r_p(t') \not\in \mathcal{B},  t_{go}>t_\text{max},\cr
-\infty, \quad r_p(t') \in \mathcal{B},
\end{cases}
\end{equation}
Here we choose $\tau=15000$. Note that the prediction of obstacle collision can be conducted at any time $t'\in[t,t_{go}]$. Here we make a short-term collision prediction by setting $t'=t+2(m+1)\varepsilon$, i.e., twice the length of a learning period.

\subsection{Results for Perfect Observations}
\begin{figure}[h]
\centering
\includegraphics[width=0.4\textwidth]{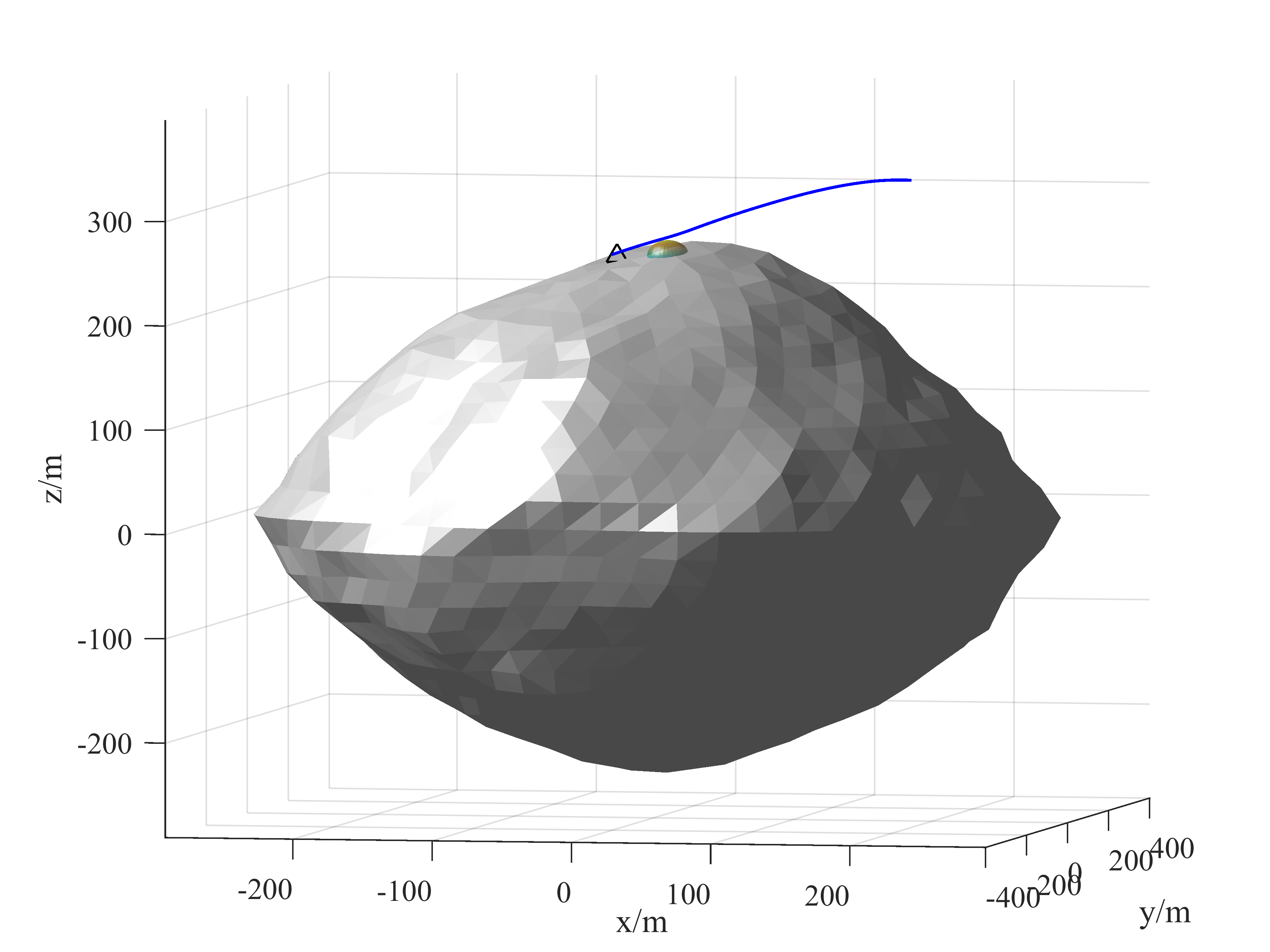}
\includegraphics[width=0.4\textwidth]{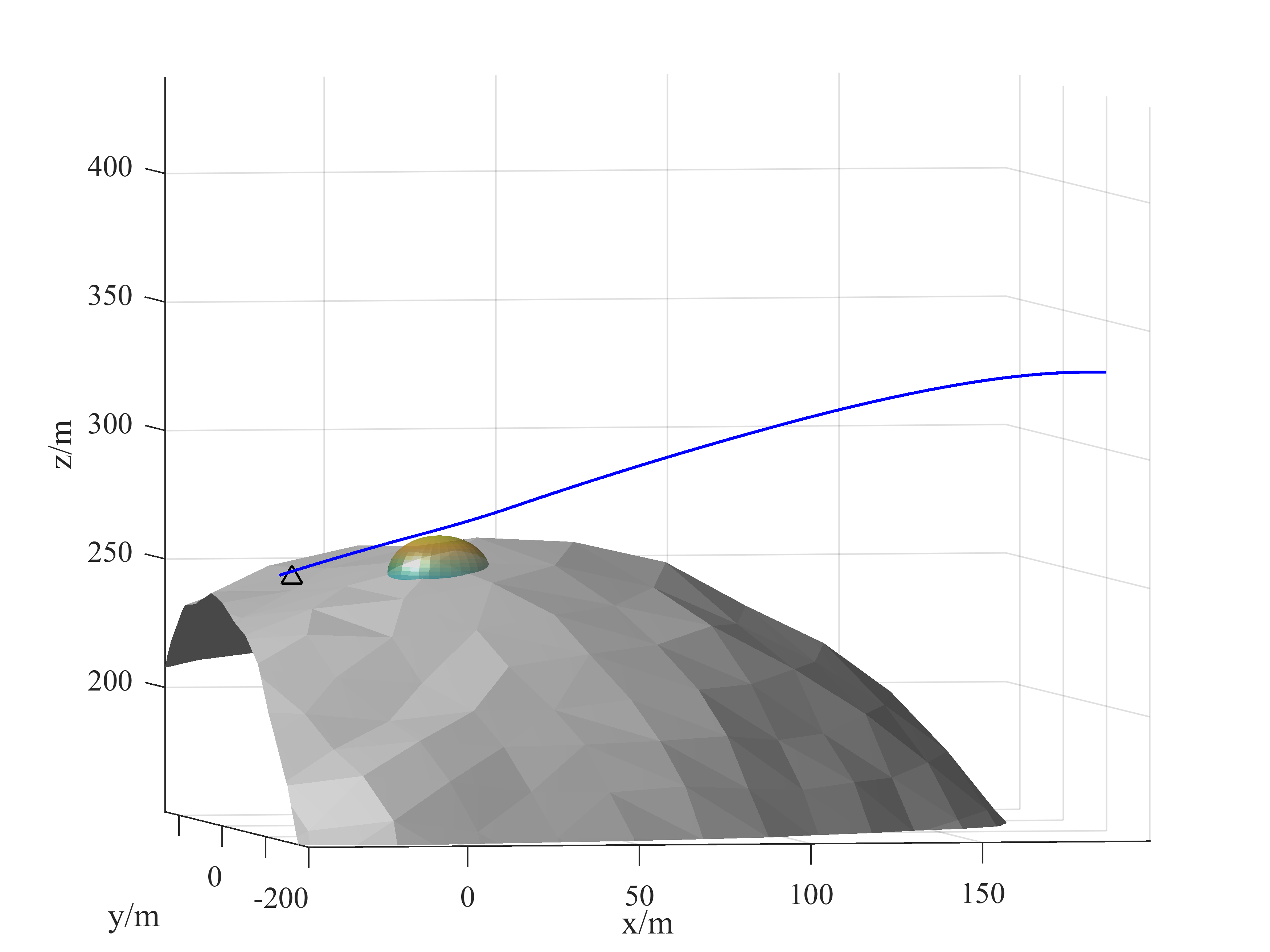}
\caption{\label{fig:preciseobs} Landing trajectory and system performance under perfect observations.}
\end{figure}

We first assume that the state observations are perfect, and apply myopic control with the goodness function defined in \eqref{defgood}. We set the vehicle initial position as $r_0=[200, -100, 330]^T\text{m}$, where the origin of the coordinate system corresponds to the asteroid center of mass. The vehicle is originally placed about $90$m above the asteroid surface. We require it to land on the surface at $r_f=[-26, 0, 243]^T\text{m}$. The obstacle sphere is centered at  $r_\mathcal{B}=[14,-23,250]^T\text{m}$ with radius $R$ of $15\text{m}$. The asteroid with the additional obstacle, as well as the starting and landing point, are illustrated in Figure \ref{fig:preciseobs}. In the simulation, the time interval between state observations is $\varepsilon=2\text{s}$. Since we apply four affine-independent control inputs $u^*+\Delta u^j$, $j=0,\ldots,3$ in each iteration, the update cycle of the local dynamics model is $(m+1)\varepsilon=8\text{s}$. The generated landing trajectory, with a $7\text{m}$ landing error, is shown in Figure \ref{fig:preciseobs}. 

\subsection{Results for Imperfect Observation}
In the asteroid landing process, the data collected by sensors is inevitably contaminated due to environment noises and instrumental calibration deviation \cite{lindner2010time}. In our simulation, we add different position observation errors on the $z$-axis, and compare system performances under nominal myopic control and robust myopic control. 

\begin{figure}[h]
\centering
\includegraphics[width=0.4\textwidth]{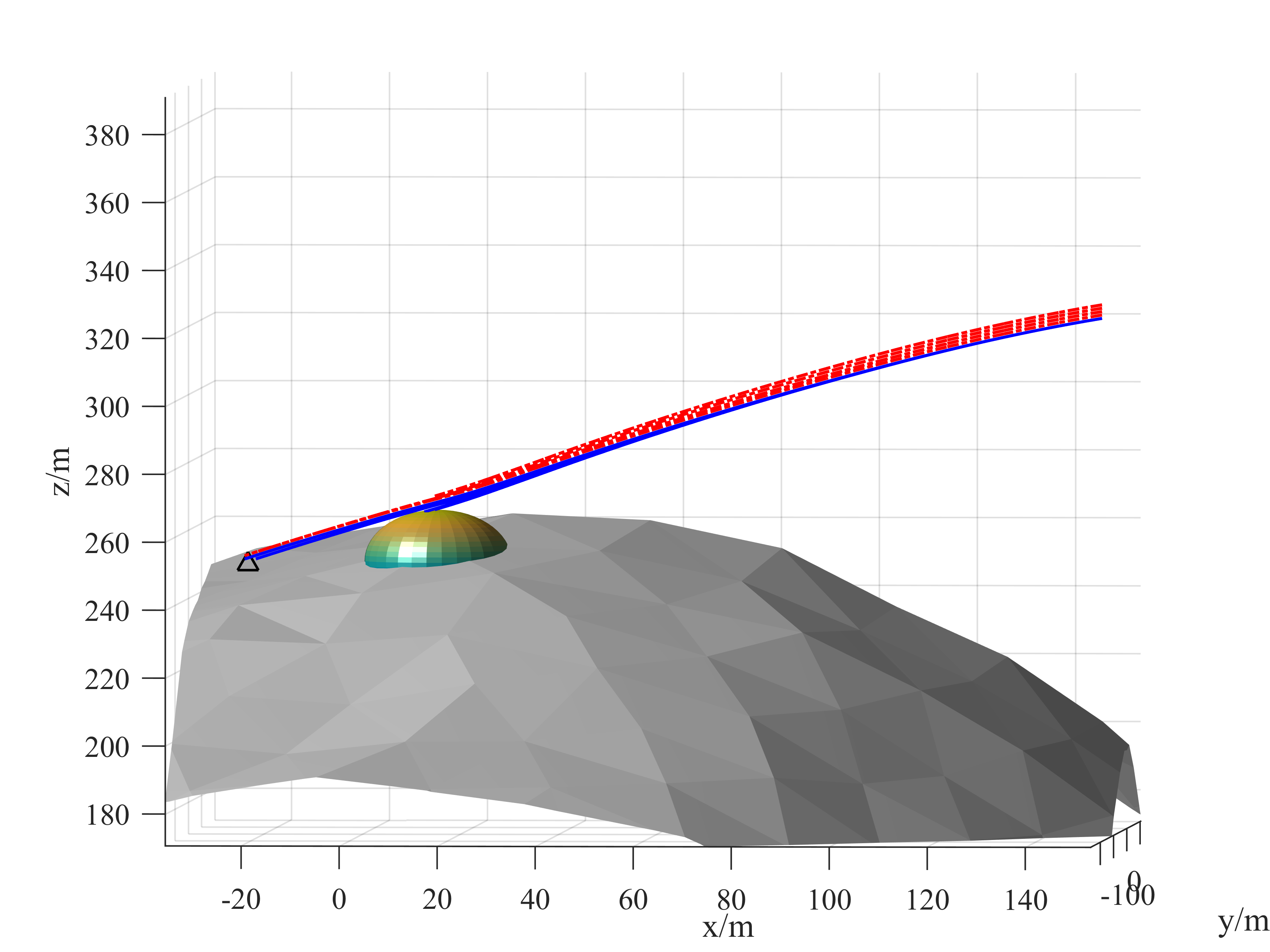}
\includegraphics[width=0.4\textwidth]{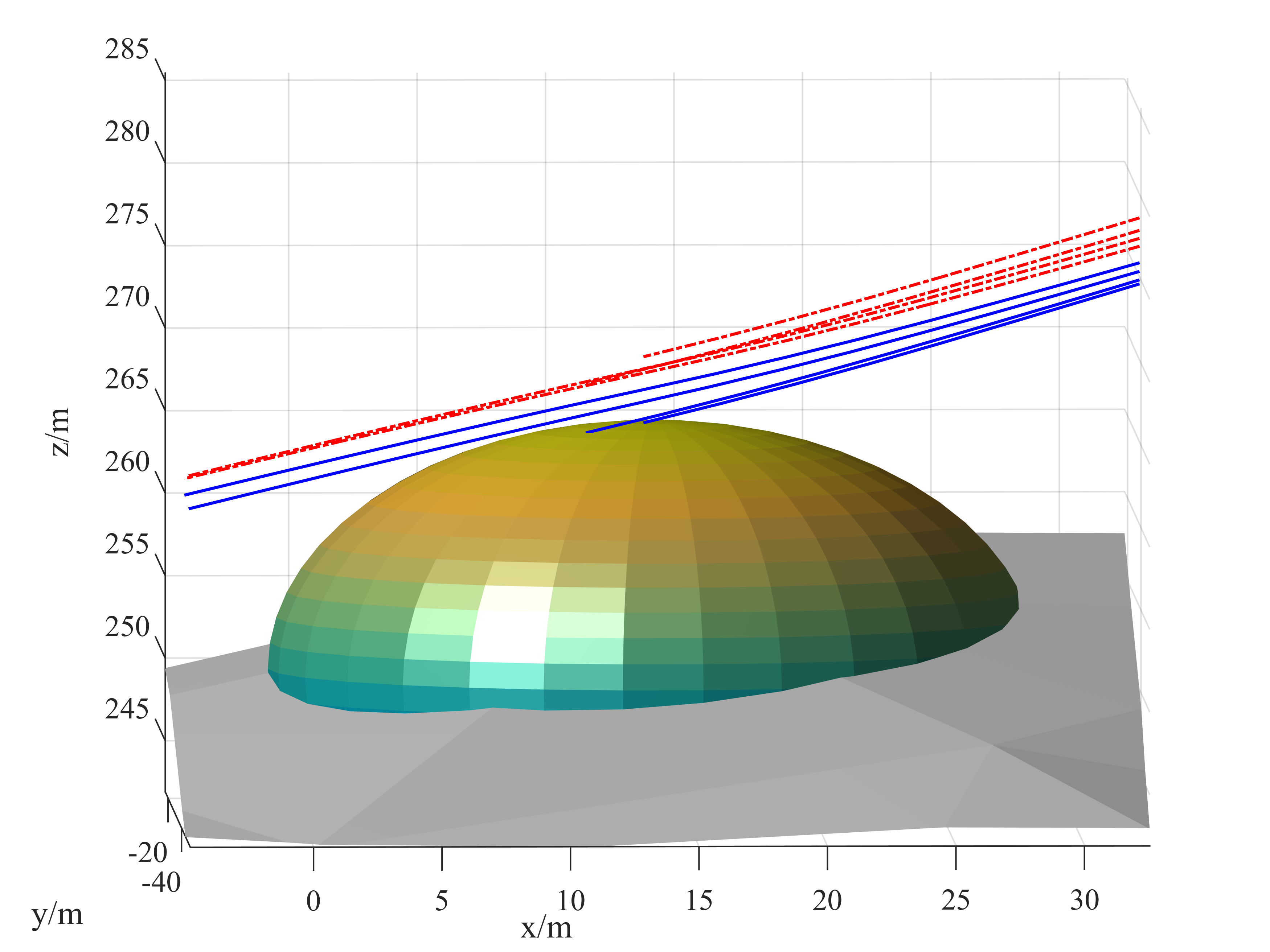}
\caption{\label{fig:NMC_IO} Actual landing trajectories (blue solid line) and observed trajectories (red dashed line) using nominal myopic control. When $e=3\text{m}$ and $e=4\text{m}$, the corresponding trajectories terminate prematurely due to intersection with the obstacle.}
\end{figure}

The nominal myopic control strategy directly uses the wrong state information in learning the dynamics and generating control actions. When the flight path is close to the obstacle in the landing process, using nominal myopic control under imperfect observation may lead to collision. Figure \ref{fig:NMC_IO} illustrates the actual and observed trajectories under observation error $e=[1,2,3,4]\text{m}$ respectively. As the error increases, the risk of intersecting with the obstacle region grows. When $e\geq3\text{m}$, the lander collides with the obstacle.

\begin{figure}[h]
\centering
\includegraphics[width=0.4\textwidth]{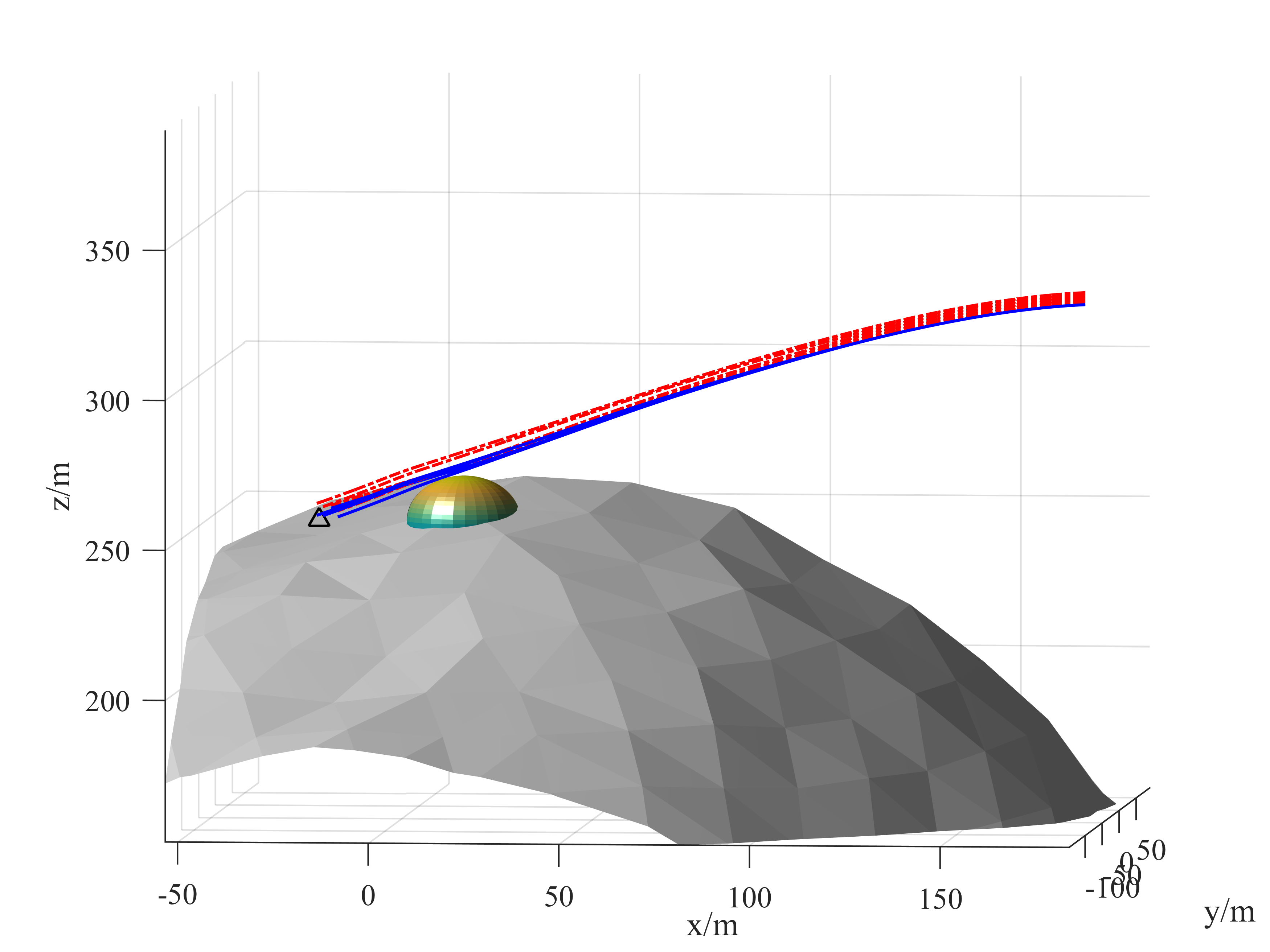}
\includegraphics[width=0.4\textwidth]{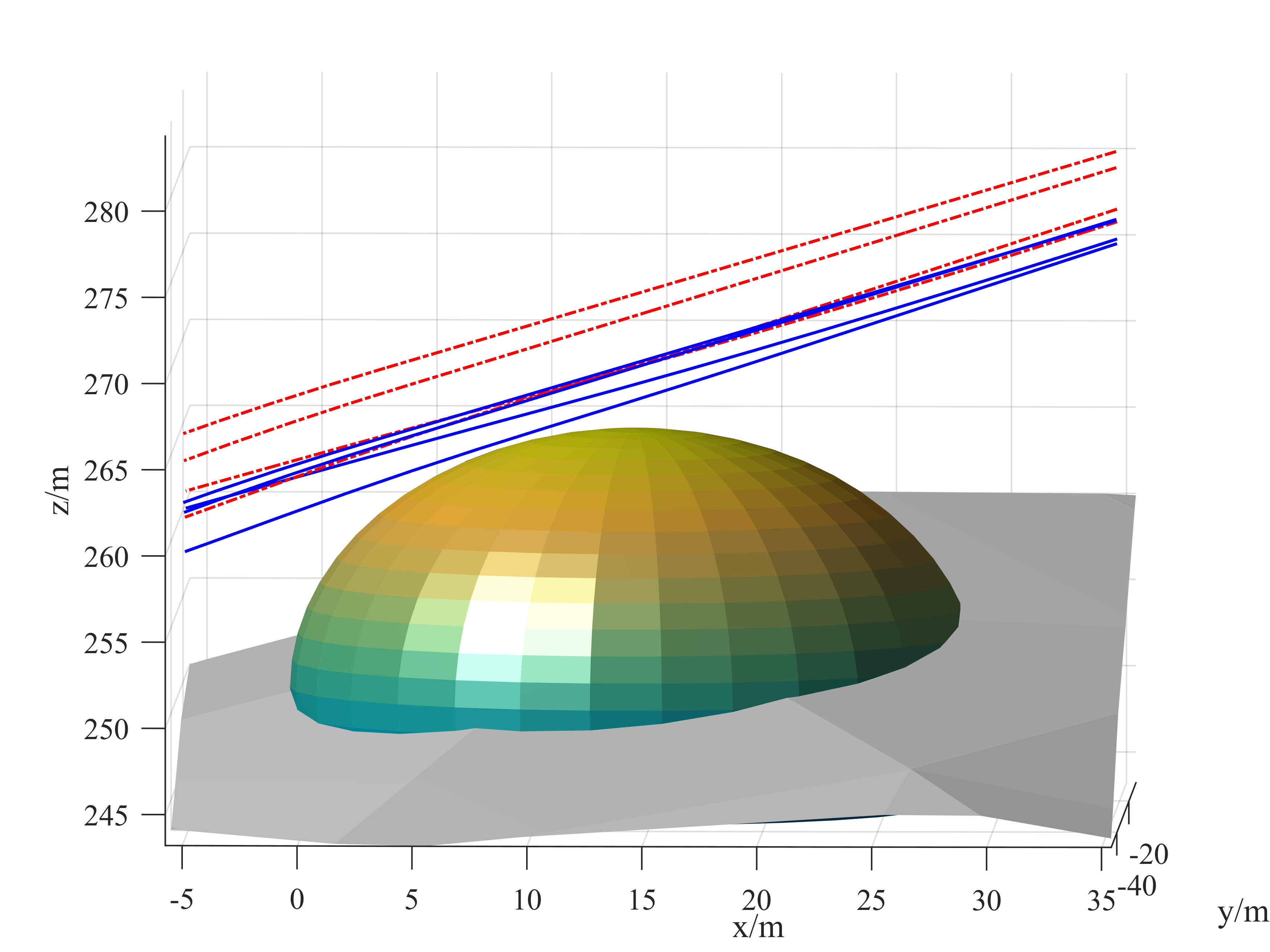}
\caption{\label{fig:RMC_IO} Actual landing trajectories (blue solid line) and observed trajectories (red dashed line) using robust myopic control. All the trajectories end at the designated landing site without collision with the obstacle.}
\end{figure}

On the other hand, the robust myopic control approach assumes that the bound of observation error $\Delta$ is known by the system in advance. In every iteration, the method generates a control action that ensures system performance even in the worst-case scenario. Here we demonstrate the results of setting $\Delta=2e$ where observation error again equals $e=[1,2,3,4]\text{m}$. The obtained trajectories are shown in Figure \ref{fig:RMC_IO}. We can see that, unlike in Figure \ref{fig:NMC_IO}, all the trajectories manage to fly over the obstacle region and reach the target.

We reiterate that, if $\Delta/e\geq 1$ and the goodness function is set to $-\infty$ for a control input whenever it is possible to result in constraint violation, then the system is guaranteed to satisfy hard constraints. However, as the value of $\Delta/e$ increases, the robust myopic control strategy becomes more conservative and computationally intractable, since a wider range of trajectories will be taken into account in decision-making. Thus, it is naturally desirable for $\Delta/e$ to be as close to $1$ as possible. In practice, the ratio $\Delta/e$ depends on the quality of our prior knowledge of the sensing uncertainties.

%% file: Conclusion.tex
In this paper, we extend the work of myopic control by considering hard constraints and imperfect observations in the problem. First, we provide guarantees for hard safety constraint satisfaction in myopic control by introducing $-\infty$ as a possible goodness function value. After that, we develop a robust myopic control strategy to deal with imperfect state observations. Given a fixed observation error bound, the method explores the set of all possible trajectories and finds an optimal control action that leads to the best-worst case performance. We then apply the method in an asteroid landing scenario based on the OSIRIS-REx mission. The result shows that when the system is moving in the vicinity of the hard constraint boundary, large observation errors will lead to constraint violation using the nominal myopic control strategy. However, by considering all possible trajectories in determining the control action, the robust myopic control approach diminishes the risk of violating constraints and thus improves system safety.

%% file: Appendix.tex
\begin{proof}
Let $t=k(m+1)\varepsilon$. In the following, we use $\phi^+$ to denote the worst case trajectory in $\mathbb{B}(\phi^\text{obs},\Delta)$ in the sense of \eqref{dandan}. Again, we abuse notation and denote by $u^+$ both the control law obtained by the robust myopic control, and its particular value at time $t=k(m+1)\varepsilon$; the use is clear from the context. In the remainder of the text, we omit the time variable: $x^\text{true}_j=\phi^{\text{true}}_{u^+}(t_0+j\varepsilon)$, $x^+_j=\phi^{+}_{u^+}(t_0+j\varepsilon)$, where $t_0=(k-1)(m+1)\varepsilon$, and all trajectories are defined on $[0,t]$.

We note that
\begin{equation}
\label{summands}
\begin{split}
& |G(\phi^\text{true}_{u^+}, v^\text{true}(u^+,x^\text{true}))-\max_{u\in\mathcal{U}} G(\phi^\text{true}_{u^+}, v^\text{true}(u,x^\text{true}))| \\ 
& \leq |G(\phi^\text{true}_{u^+}, v^\text{true}(u^+,x^\text{true}))-G(\phi^\text{true}_{u^+}, v^\text{true}(u^*,x^\text{true}))| \\
& + |G(\phi^\text{true}_{u^+}, v^\text{true}(u^*,x^\text{true}))-\max_{u\in\mathcal{U}} G(\phi^\text{true}_{u^+}, v^\text{true}(u,x^\text{true}))|
\end{split}
\end{equation}
by triangle inequality, where $u^*$, given by \eqref{dantong2}, is the myopically optimal control based on learned dynamics for trajectory $\phi^\text{true}_{u^+}$. The last summand in \eqref{summands} would be equal to $0$ if $v=v^\text{true}$, i.e., if the system perfectly learned the dynamics. However, since $v\neq v^\text{true}$, this summand can be bounded from Theorem 13 in \cite{Ornik} by 
\begin{equation}
\label{newbo}
4LM_0M_1(m + 1)^3(4m^{\frac{3}{2}}+\delta)\frac{\varepsilon}{\delta}.
\end{equation}

Now, based on the triangle inequality, 
\begin{align}
\label{rob11}
\begin{split}
&|G(\phi^\text{true}_{u^+}, v^\text{true}(u^+,x^\text{true}))-G(\phi^\text{true}_{u^+}, v^\text{true}(u^*,x^\text{true}))|\\
& \leq |G(\phi^\text{true}_{u^+}, v^\text{true}(u^+,x^\text{true}))-G(\phi^+_{u^+}, v(u^+,x^+))| \\
&+|G(\phi^+_{u^+}, v(u^+,x^+))-G(\phi^\text{true}_{u^+}, v^\text{true}(u^*,x^\text{true}))|,
\end{split}
\end{align}
where, through an understandable abuse of notation, $v$ is obtained from learning using the corresponding trajectory: e.g., $v(u^+,x^+)=\sum\lambda_j^+(x_{j+1}^+-x_j^+)/\varepsilon$.
Define
\begin{align}
\Vert \phi^\text{true}_{u^+}-\phi^+_{u^+}\Vert \doteq \max_{\substack{x^\text{true}_j \in \phi^\text{true}_{u^+},\\ x^+_j \in \phi^+_{u^+},\\j\in[m]}} \Vert x^\text{true}_j-x^+_j\|.
\end{align}
The dynamics learned from $x^+$ approximate true dynamics when $x^\text{true}$ and $x^+$ are close. Define $\Delta'_j=x^+_j-x^\text{true}_j$ for all $j\in [m]$. Since the goodness function $G$ has Lipschitz constant $L$, the first part on the right hand side of \eqref{rob11} can be bounded by
\begin{align}
\label{firstpart}
\begin{split}
&|G(\phi^\text{true}_{u^+}, v^\text{true}(u^+,x^\text{true}))-G(\phi^+_{u^+}, v(u^+,x^+))|\\
& \leq L(\Vert \phi^\text{true}_{u^+}-\phi^+_{u^+}\Vert +\Vert v^\text{true}(u^+, x^\text{true})-v(u^+, x^+)\Vert )\\
& \leq L(\max_{j\in [m]}\Vert x^\text{true}_j-x^+_j\Vert+\Vert v^\text{true}(u^+, x^\text{true})-v(u^+, x^+)\Vert)\\
& \leq L(\max_{j\in [m]}\Vert \Delta'_j\Vert+\Vert v^\text{true}(u^+, x^\text{true})-v(u^+, x^+)\Vert).
\end{split}
\end{align}
We now bound $\Vert v^\text{true}(u^+, x^\text{true})-v(u^+, x^+)\Vert$:
\begin{align} 
\label{dynerror}
\begin{split}
&\Vert v^\text{true}(u^+, x^\text{true})-v(u^+, x^+)\Vert = \left\Vert (f(x) + \sum g_i u^+_i) - \sum^{m}_{j=0} \lambda_j ^+\frac{x^+_{j+1} - x^+_j}{\varepsilon} \right\Vert \\
&= \left\Vert (f(x) + \sum g_i u^+_i) - \sum^{m}_{j=0} \lambda_j ^+\frac{x^\text{true}_{j+1} + \Delta'_{j+1} - x^\text{true}_j - \Delta'_j}{\varepsilon}\right\Vert \\
&\leq \left\Vert (f(x) + \sum g_i u^+_i) - \sum^{m}_{j=0} \lambda_j ^+\frac{x^\text{true}_{j+1}- x^\text{true}_j}{\varepsilon} \right\Vert + \left\Vert \sum_{j=0}^m \lambda_j ^+\frac{\Delta'_{j+1} - \Delta'_{j}}{\varepsilon} \right\Vert\\
& \leq \left\Vert (f(x) + \sum g_i u^+_i) - \sum^{m}_{j=0} \lambda_j ^+\frac{x^\text{true}_{j+1}- x^\text{true}_j}{\varepsilon}\right\Vert + \frac{2\max_{j\in[m]}\|\Delta'_{j}\|}{\varepsilon}\left( 1 + 4m \frac{\sqrt{m}}{\delta} \right)\\
&\leq  2 M_0 M_1 (m+1)^3 (4 m^{\frac{3}{2}} + \delta) \frac{\varepsilon}{\delta}+ \frac{2\max_{j\in[m]}\|\Delta'_{j}\|}{\varepsilon}\left( 1 + 4m \frac{\sqrt{m}}{\delta} \right),
\end{split}
\end{align}
where the bound for $\Vert\sum_{j=0}^m \lambda_j ^+ \Vert$ together with the dynamics learning error bound are proved in \cite{Ornik}. Combining \eqref{firstpart} and \eqref{dynerror}, we find the bound for the first part on the right hand of \eqref{rob11} as
\begin{align}
&|G(\phi^\text{true}_{u^+}, v^\text{true}(u^+,x^\text{true}))-G(\phi^+_{u^+}, v(u^+,x^+))| \nonumber\\
&\leq 2L M_0 M_1 (m+1)^3 (4 m^{\frac{3}{2}} + \delta) \frac{\varepsilon}{\delta}+L\left( \max_{j\in [m]} \Vert \Delta'_j\Vert + \frac{2\max_{j\in[m]}\|\Delta'_{j}\|}{\varepsilon}( 1 + 4m \frac{\sqrt{m}}{\delta} )\right) \label{part1}
\end{align}

Let us now bound $|G(\phi^+_{u^+}, v(u^+,x^+))-G(\phi^\text{true}_{u^+}, v^\text{true}(u^*,x^\text{true}))|$. Applying the optimal control $u^*$ generated from perfect state observations on all possible true trajectories, there exists a trajectory that receives the minimum goodness function value. Denote this trajectory and the states on it as $\phi^0_{u^*}$ and $x^0$. Then, we have
\begin{align}
G(\phi^+_{u^+},v(u^+, x^+)) \geq G(\phi^0_{u^+}, v(u^*, x^0)).
\end{align}
Considering that
\begin{align} 
\begin{split}
&|G(\phi^0_{u^+}, v(u^*, x^0))-G(\phi^\text{true}_{u^+}, v^\text{true}(u^*,x^\text{true}))| \\
&\leq |G(\phi^0_{u^+}, v(u^*, x^0))-G(\phi^\text{true}_{u^+}, v(u^*,x^\text{true}))|\\
&+ |G(\phi^\text{true}_{u^+}, v(u^*,x^\text{true}))-G(\phi^\text{true}_{u^+}, v^\text{true}(u^*,x^\text{true}))| \\
&\leq L(2\max_{j\in[m]}\|\Delta_j'\| + \Vert v(u^*, x^0)- v(u^*,x^\text{true}) \Vert) + 2L M_0 M_1 (m+1)^3 (4 m^{\frac{3}{2}} + \delta) \frac{\varepsilon}{\delta},
\end{split}
\end{align} 
and
\begin{align} 
\begin{split}
&\Vert v(u^*, x^0)- v(u^*,x^\text{true}) \Vert\\
&= \left\Vert \sum^{m}_{j=0}\lambda_j \frac{x^0_{j+1} - x^0_j}{\varepsilon} - \sum^{m}_{j=0} \lambda_j \frac{x^\text{true}_{j+1} - x^\text{true}_j}{\varepsilon} \right\Vert  \\
&= \left\Vert \sum^{m}_{j=0}\lambda_j \frac{(x^0_{j+1} -x^\text{true}_{j+1}) - (x^0_j - x^\text{true}_j)}{\varepsilon}\right\Vert \\
&\leq \frac{2\max_{j\in[m]}\|\Delta_j'\|}{\varepsilon}\left( 1 + 4m \frac{\sqrt{m}}{\delta}\right),
\end{split}
\end{align}
we obtain
\begin{align}
\label{ub}
\begin{split}
&G(\phi^\text{true}_{u^+}, v^\text{true}(u^*,x^\text{true})) - G(\phi^+_{u^+},v(u^+, x^+)) \\
& \leq G(\phi^\text{true}_{u^+}, v^\text{true}(u^*,x^\text{true})) - G(\phi^0_{u^+}, v(u^*, x^0))\\
& \leq L\left(2\max_{j\in[m]}\|\Delta_j'\| + \frac{2\max_{j\in[m]}\|\Delta_j'\|}{\varepsilon}( 1 + 4m \frac{\sqrt{m}}{\delta})\right) + 2L M_0 M_1 (m+1)^3 (4 m^{\frac{3}{2}} + \delta) \frac{\varepsilon}{\delta}.
\end{split}
\end{align}
On the other hand, as $\phi^+_{u^+}$ is the worst case trajectory under control $u^+$, and $u^*$ is the myopically optimal control for trajectory $\phi^\text{true}_{u^+}$, we have
\begin{align}
G(\phi^+_{u^+},v(u^+, x^+)) \leq G(\phi^\text{true}_{u^+},v(u^+, x^\text{true}))\leq G(\phi^\text{true}_{u^+},v(u^*, x^\text{true})).
\end{align}
Hence, $G(\phi^\text{true}_{u^+}, v^\text{true}(u^*,x^\text{true})) - G(\phi^+_{u^+},v(u^+, x^+))$ can be bounded from below:
\begin{align}
\label{lb}
\begin{split}
&G(\phi^\text{true}_{u^+}, v^\text{true}(u^*,x^\text{true})) - G(\phi^+_{u^+},v(u^+, x^+)) \\
& \geq G(\phi^\text{true}_{u^+}, v^\text{true}(u^*,x^\text{true})) - G(\phi^\text{true}_{u^+},v(u^+, x^\text{true}))\\
& \geq G(\phi^\text{true}_{u^+}, v^\text{true}(u^*,x^\text{true})) - G(\phi^\text{true}_{u^+},v(u^*, x^\text{true}))\\
& \geq -2L M_0 M_1 (m+1)^3 (4 m^{\frac{3}{2}} + \delta) \frac{\varepsilon}{\delta}.
\end{split}
\end{align}
From \eqref{ub} and \eqref{lb}, we obtain
\begin{align}
&|G(\phi^+_{u^+}, v(u^+,x^+))-G(\phi^\text{true}_{u^+}, v^\text{true}(u^*,x^\text{true}))|\nonumber\\
& \leq L\left(2\max_{j\in[m]}\|\Delta_j'\| + \frac{2\max_{j\in[m]}\|\Delta_j'\|}{\varepsilon}( 1 + 4m \frac{\sqrt{m}}{\delta})\right) + 2L M_0 M_1 (m+1)^3 (4 m^{\frac{3}{2}} + \delta) \frac{\varepsilon}{\delta} \label{part2}
\end{align}
Combining \eqref{newbo}, \eqref{rob11}, \eqref{part1}, and \eqref{part2}, we obtain
\begin{align}
\begin{split}
&|G(\phi^\text{true}_{u^+}, v^\text{true}(u^+,x^\text{true}))-\max_{u\in\mathcal{U}} G(\phi^\text{true}_{u^+}, v^\text{true}(u,x^\text{true}))| \\
&\leq 8L M_0 M_1 (m+1)^3 (4 m^{\frac{3}{2}} + \delta) \frac{\varepsilon}{\delta}+L\left(3\max_{j\in [m]} \Vert \Delta'_j\Vert+\frac{4\max_{j\in[m]}\|\Delta_j'\|}{\varepsilon}\left( 1 + 4m \frac{\sqrt{m}}{\delta}\right)\right)\\
&\leq 8L M_0 M_1 (m+1)^3 (4 m^{\frac{3}{2}} + \delta) \frac{\varepsilon}{\delta}+2L\left(3\Delta+\frac{4\Delta}{\varepsilon}\left( 1 + 4m \frac{\sqrt{m}}{\delta}\right)\right).
\end{split}
\end{align}
\end{proof}